# Multiple supersonic phase fronts launched at a complex-oxide hetero-interface

M. Först[1,2*], K.R. Beyerlein[2,3], R. Mankowsky[1,2], W. Hu[1,2], G. Mattoni[4], S. Catalano[5], M. Gibert[5], O. Yefanov[2,3], J.N. Clark[2,6], A. Frano[7,8], J.M. Glownia[9], M. Chollet[9], H. Lemke[9], B. Moser[10], S.P. Collins[10], S.S. Dhesi[10], A.D. Caviglia[4], J.-M. Triscone[5], and A. Cavalleri[1,2,11#]

[1]Max Planck Institute for the Structure and Dynamics of Matter, 22761 Hamburg, Germany
[2]Center for Free Electron Laser Science, 22761 Hamburg, Germany
[3]Deutsches Elektronen-Synchrotron DESY, 22607 Hamburg, Germany
[4]Kavli Institute of Nanoscience, Delft University of Technology, 2628 CJ Delft, The Netherlands
[5]Department of Quantum Matter Physics, Université de Genève, 1211 Genève, Switzerland
[6]Stanford Pulse Institute, SLAC National Accelerator Laboratory, Menlo Park 94025, California, USA
[7]Materials Sciences Division, Lawrence Berkeley National Laboratory, Berkeley 94720, California, USA
[8]Advanced Light Source, Lawrence Berkeley National Laboratory, Berkeley 94720, California, USA
[9]Linac Coherent Light Source, SLAC National Accelerator Laboratory, Menlo Park 94025, California, USA
[10]Diamond Light Source, Didcot, Oxfordshire OX11 0QX, UK
[11]Department of Physics, Clarendon Laboratory, University of Oxford, Oxford OX1 3PU, United Kingdom

* michael.foerst@mpsd.mpg.de; # andrea.cavalleri@mpsd.mpg.de

## Abstract

Selective optical excitation of a substrate lattice can drive phase changes across hetero-interfaces. This phenomenon is a non-equilibrium analogue of static strain control in heterostructures and may lead to new applications in optically controlled phase change devices. Here, we make use of time-resolved non-resonant and resonant x-ray diffraction to clarify the underlying physics, and to separate different microscopic degrees of freedom in space and time. We measure the dynamics of the lattice and that of the charge disproportionation in $NdNiO_3$, when an insulator-metal transition is driven by coherent lattice distortions in the $LaAlO_3$ substrate. We find that charge redistribution propagates at supersonic speeds from the interface into the $NdNiO_3$ film, followed by a sonic lattice wave. When combined with measurements of magnetic disordering and of the metal-insulator transition, these results establish a hierarchy of events for ultrafast control at complex oxide hetero-interfaces.

Complex oxide heterostructures have emerged as a versatile means to engineer functional materials at equilibrium [1,2,3]. In nickelate thin films, for example, electronic and magnetic properties have been controlled statically by producing interfacial strain through the substrate [4,5,6].

Insulator-metal, magnetic and structural transitions can also be induced dynamically by controlling these hetero-interfaces with light, particularly when the crystal lattice of the substrate is deformed coherently with mid-infrared radiation. In the case of $LaAlO_3$/$NdNiO_3$ heterostructures, a long-lived metallic phase can be triggered in the $NdNiO_3$ film by exciting large amplitude vibrations in the $LaAlO_3$ substrate [7]. This new type of optically controlled phase transition is very appealing as a basis for fast phase change devices, in which the optical control is spatially separated from the functional material. One can envisage device architectures, in which phase fronts transport and manipulate information in new ways.

The underlying physics and the role of different orders in space and time has been investigated in some detail in recent experiments. Ultrafast resonant soft x-ray diffraction experiments have, for example, revealed that a supersonic paramagnetic/antiferromagnetic front accompanies the insulator metal transition (IMT), propagating from the hetero-interface into the material [8]. Additional insight was provided by optical experiments in a related $LaAlO_3$/$SmNiO_3$ heterostructure, in which the same substrate excitation was shown to induce an IMT in the nickelate film also above the Néel temperature [9]. This observation was taken as evidence that melting of magnetic order may follow the IMT and may not be its cause. Loss of charge disproportionation alone, launched at the interface and propagating into the material, may in fact be the root cause for the IMT.

Here, we complete this physical picture with femtosecond non-resonant and resonant *hard* x-ray diffraction (XRD) experiments, which are used to measure the rearrangement of the lattice and of the ordered charges. These data, combined with previous evidence, show that different types of order-disorder and structural phase fronts are found to propagate at different speeds into the functional material, with charge order melting presumably being the driving force.

Thin films of $NdNiO_3$ were grown on twinned (111) $LaAlO_3$ single crystals [10,11], reducing the metal-insulator transition temperature, observed at ~200 K in the bulk [12], to ~130 K. Across the bulk phase transition, the orthorhombic (*Pbnm*) crystal structure of the metallic state transforms into a monoclinic (*P21/n*) structure [13] with two alternating $NiO_6$ octahedra of expanded and contracted Ni-O bonds [14] that is accompanied by antiferromagnetic ordering [15] and charge disproportionation [16].

The equilibrium lattice rearrangements were first studied across the temperature driven phase transition using 8-keV x-rays from a synchrotron light source (Diamond Light Source, I16 beamline). The inset of Figure 1(a) shows the orthorhombic $(055)_{Pbnm}$ diffraction peak – equivalent to the pseudo-cubic $(2½\ \overline{2½}\ 2½)$ reflection along the sample out-of-plane direction – with a clear shift in momentum below and above the transition temperature. This shift in the $(055)_{Pbnm}$ peak position as a function of temperature is plotted in the main panel, establishing the hysteretic structural phase transition that accompanies the equilibrium electronic and magnetic changes. The observed ~0.15% decrease in wave vector $q$ results from the discontinuity in unit-cell volume associated with this first-order phase transition [13]. Note, however, that the in-plane lattice parameter of the thin film is clamped by the substrate so that only the projection of the orthorhombic $b$ and $c$ axes on the out-of-plane direction contribute to the observed contraction [17].

Time-resolved experiments were performed at an x-ray free electron laser (Linac Coherent Light Source at the SLAC Laboratory) to measure the dynamic lattice rearrangements. The ~13 nm nickelate thin film was cooled by a cryogenic Helium gas jet into the antiferromagnetic insulating state to approximately 40 K. The sample was excited by laser pulses 200-fs in duration and 3.6-mJ/cm² of fluence at 15 µm wavelength, resonant with the highest-frequency $LaAlO_3$ substrate phonon (see Fig. 1(b), right panel). The x-ray probe pulses (<100 fs duration) were tuned to 8.0 keV photon energy and spectrally filtered to ~1 eV by a channel-cut Si(111) monochromator. The diffracted x-rays were detected by a two-dimensional pixel array (Cornell-SLAC hybrid Pixel Array Detector, CSPAD), integrated over the detector area, and normalized on a shot-to-shot basis to the incident x-ray intensity.

Figure 1(b) shows the dynamical changes in the $NdNiO_3$ orthorhombic $(505)_{Pbnm}$ reflection, equivalent to the (2½ 2½ 2½) diffraction peak in the pseudo-cubic notation. The peak intensity was observed to reduce by ~50 % on the time scale of about 3 ps, well resolved by the instrument's temporal resolution of 250 fs, before recovering with a double-exponential profile and time constants of 40 and 300 ps. Concomitantly, shifts to lower $q$ in the diffraction peak evidence lattice expansion along the orthorhombic $(101)_{Pbnm}$ out-of-plane direction.

Figure 1(c) displays the $(505)_{Pbnm}$ and $(055)_{Pbnm}$ diffraction peaks, accessible by measuring the NdNiO$_3$ thin film at different domains of the twinned substrate, at +4 ps time delay and at equilibrium. The $(505)_{Pbnm}$ reflection shifts to lower $q$ by about −0.15%, whilst the $(055)_{Pbnm}$ reflection shifts in the opposite direction by +0.05%. The sign and relative magnitude of these dynamical shifts reproduce well those observed for the equilibrium IMT of bulk NdNiO$_3$, which involves expansion along the *a* axis and contraction along the *b* axis by different amounts [13]. Also, the light-induced relative intensity change of the (055) peak is commensurate with the equilibrium phase transition, plotted in the inset of Fig.1(a).

We assume that this structural transition propagates from the hetero-interface into the film, as discussed in Reference 8. This assessment is based on the fact that the insulator-metal response is only observed when the pump field is made resonant with the phonon of the substrate [7]. Hence, the effect must be associated with dynamics occurring at the buried interface.

The speed of the corresponding phase front can be extracted from the transient changes of the $(505)_{Pbnm}$ peak position. We consider that the ~13 nm NdNiO$_3$ thin film can, at each time delay, be divided into two layers of thicknesses $d_1$ and $d_2$, whose out-of-plane lattice spacing are either that of the equilibrium insulating film (top nickelate layer) or that of the light-induced state (bottom layer next to the interface). The measured strain $\varepsilon = \Delta q/q_0$ descends as the weighted sum of the strain values present in each layer [18]

$$\varepsilon = \frac{\varepsilon_1 d_1 + \varepsilon_2 d_2}{d_1 + d_2} . \tag{1}$$

The transient strain was determined by fitting Gaussian functions to the diffraction peaks at each time delay. Assuming that the whole film has been transformed at +4 ps delay, we find that a structural transformation towards the high-temperature equilibrium orthorhombic *Pbnm* lattice propagates at about a speed of 4 nm/ps (see Figure 4), which matches very closely with the known longitudinal sound velocity in NdNiO$_3$ [19].

We next turn to the dynamical changes in charge disproportionation, which are accessible by tuning the hard x-ray pulses into resonance with the 8.34-keV Ni K-edge. For NdNiO$_3$, the (0$kl$)$_{Pbnm}$ and ($h$0$l$)$_{Pbnm}$ reflections (with $h$, $k$ and $l$ being odd indices) are particularly sensitive to the ordering on the two inequivalent Ni sites, and show resonant enhancement in the diffraction intensity [20,21]. Figure 2(a) displays the associated intensity modulation of the (505)$_{Pbnm}$ reflection as a function of the LCLS photon energy in the vicinity of the Ni K-edge (T=40 K, blue data points). These data agree well with a more comprehensive demonstration of this effect, measured with synchrotron radiation at the NdNiO$_3$ (101)$_{Pbnm}$ reflection (black solid line, this data is taken from Ref. 22).

Figure 2(b) presents the dynamical changes of the (505)$_{Pbnm}$ reflection, comparing experiments with x-ray photons resonant (8.346 keV) and non-resonant (8.329 keV) with the Ni K-edge. The non-resonant probe yields a reduction in peak intensity within 3 ps, as observed for 8-keV radiation (see Fig.1). At negative time delays, the peak intensity for resonant diffraction exhibits an additional contribution arising from the charge disproportionation (illustrated by the grey shaded region), which rapidly disappears after the optical excitation and hence leads to the convergence of the two intensities. Within good approximation, the corresponding order parameter is proportional to $\sqrt{I_{on}} - \sqrt{I_{off}}$, with $I_{on/off}$ being the resonant and off-resonant peak intensities [20]. This quantity is plotted in Fig. 2(c), showing that the substrate phonon excitation leads to complete loss of charge disproportionation in the NdNiO$_3$ film within less than 1.5 ps (blue data points). Under the assumption of a melt front propagating from the interface across the whole film, the linear slope in the error function fit to these dynamics (red solid line) corresponds to a front velocity for the charge order melting of ~11 nm/ps, more than twice the NdNiO$_3$ speed of sound and far faster than the lattice rearrangement observed above.

Presumably, the loss of charge disproportionation defines the speed, at which the insulator-metal transition takes place. This is confirmed by time-resolved THz spectroscopy experiments, which were performed for the same excitation of the LaAlO$_3$ substrate phonon. The transient low-frequency reflectivity of the heterostructure was measured by THz pulses generated in a laser-ionized gas plasma [23] and detected between 50 and 250 cm$^{-1}$ via electro-optic sampling in a GaP crystal. The spectra for a ~33 nm NdNiO$_3$ film (measured at 10 K) are shown in Figure 3. The light-induced IMT is evidenced here by the pronounced increase in the Drude contribution to the low-frequency reflectivity. The transient conductivity of the nickelate reaches the equilibrium metallic value within less than 2 ps, as reported before in Refs. 7 and 9.

These time-resolved THz data were fitted using a three-layer model, and by assuming a metallic phase that propagates from the substrate into the insulating NdNiO$_3$ film [24]. Here, the nickelate film is divided into two layers, of which layer 1 represents the top part retaining the equilibrium properties and layer 2 represents the bottom part next to the interface, which is transformed from the insulating into the metallic state. Layer 3 is the LaAlO$_3$ substrate. With increasing time delay, the thickness of layer 2 increases (with layer 1 thickness simultaneously decreasing) until the whole NdNiO$_3$ film is transformed into a metallic state. The transient reflection coefficient of the heterostructure can be fitted as

$$\tilde{r}_{123} = \frac{\tilde{r}_1 + \tilde{r}_2 exp(2i\delta_2) + \tilde{r}_3[2i(\delta_2+\delta_3)] + \tilde{r}_1\tilde{r}_2\tilde{r}_3 exp(2i\delta_3)}{1 + \tilde{r}_1\tilde{r}_2 exp(2i\delta_2) + \tilde{r}_2\tilde{r}_3 exp(2i\delta_3) + \tilde{r}_1\tilde{r}_3 \exp[2i(\delta_2+\delta_3)]} \tag{1}$$

where $\delta_m = 2\pi d_m(n_m + ik_m)/\lambda$ (m = 1,2,3). Here $n_m$ and $k_m$ are the real and imaginary parts of the refractive indices, $d_m$ the layer thicknesses, and $\lambda$ is the THz probe wavelength. $\tilde{r}_{123}$ and $\tilde{r}_m$ are the electric field reflection coefficients of the whole heterostructure and of the three layers described above. Since the bare substrate does not show any light-induced changes [9] and the transient conductivity of the nickelate film reaches the equilibrium metallic value, we used the equilibrium optical properties of LaAlO$_3$ and of insulating (metallic) NdNiO$_3$ also in the transient states of the three layers [25]. The only time-varying fitting parameters for the above equation are the thicknesses $d_1$ and $d_2$ of the NdNiO$_3$ layer.

Figure 4 summarizes the different evolution of the involved phase fronts. The insulator-metal front (blue data points) determined from these THz data, appears to advance at the same speed as the charge order melt front of 11·10$^3$ m/s (blue solid line) determined above from the resonant diffraction data. These are followed by the magnetic melt front, which propagates at about 6·10$^3$ m/s (red data, taken from Ref. 8), and the structural phase front propagating at about $4 \cdot 10^3$ m/s (black data points, extracted from off-resonant x-ray diffraction presented in Fig. 1), equal to the NdNiO$_3$ speed of sound (grey solid line).

The rates, at which these individual phase fronts advance, provide a mechanistic description of the dynamical phase transition initiated at the substrate-film interface. It is also apparent from Fig. 4 that the IMT and the emergence of itinerant charges may be driving rearrangements also in the magnetic and structural properties in the nickelate film. This analysis substantiates our earlier proposal of a phase transition driven by charge redistribution [8]. Furthermore, we can safely exclude a propagating acoustic (strain) wave as a driving force for the observed dynamics. This conclusion, which was tentatively deducted from previous measurements of the magnetic front alone, is now far clearer.

The dynamics observed here also underscore the non-equilibrium nature of this optically driven phase change. In those nickelates that show a separation of the phase transition temperatures like SmNiO$_3$ or EuNiO$_3$, the Néel temperature is always lower than the metal-insulator transition temperature [26].

The data reported here shows also that a driven phase transition, that is one not initiated by thermal or quantum fluctuations near equilibrium but by external, nonlinear stimulation, can propagate at speeds that are not limited by the speed of sound. This may provide new paradigms for high-speed devices.

Finally, we note that for time delays between 1 and 3 picoseconds NdNiO$_3$ is in a metallic phase, while "inertially" retaining magnetic and structural properties of the low temperature phase (see Figure 4). This suggests that NdNiO$_3$ may be made metallic solely by the redistribution of charges, with the lattice and magnetic rearrangements being secondary order parameters.

To summarize, we have used a combination of multiple femtosecond optical and x-ray probes to provide a complete picture of the space and time dynamics of an insulator-metal transition induced by lattice excitation in a complex oxide hetero-interface. The dynamics at the interface remain unclear, and should be analyzed with sensitive probes of atomic, magnetic and electronic structures, a far more difficult set of measurements in time-resolved geometry.

## ACKNOWLEDGEMENT

Use of the Linac Coherent Light Source (LCLS), SLAC National Accelerator Laboratory, is supported by the U.S. Department of Energy, Office of Science, Office of Basic Energy Sciences under Contract No. DE-AC02-76SF00515. Also, we acknowledge the provision of beamtime by Diamond Light Source under proposal number NT-11118.

The research leading to these results has received funding from the European Research Council under the European Union's Seventh Framework Programme (FP7/2007-2013) / ERC Grant Agreement n° 319286 (Q-MAC).

**FIGURE CAPTIONS**

**Fig. 1:** (a) Temperature dependent hysteretic peak position of the $(055)_{Pbnm}$ diffraction peak, measured from the NdNiO$_3$ thin film at 8 keV photon energy. The inset shows corresponding diffraction peaks ($\theta$-$2\theta$ scans) measured below and above the phase transition temperature. Data taken at Diamond Light Source, I16 beamline. (b) Time evolution of the NdNiO$_3$ $(505)_{Pbnm}$ reflection ($\theta$-$2\theta$ scans) at 8 keV photon energy and at 40 K nominal sample temperature, induced by the direct lattice excitation of the LaAlO$_3$ substrate with 15-µm wavelength pulses of ~3.6 mJ/cm² fluence. The diffraction intensity, normalized to the incident x-ray intensity, is color-coded on a linear scale. Data taken at the LCLS free electron laser, XPP beamline. The right panel shows the mid-infrared excitation spectrum (black data points, Gaussian fit in red), compared to the resonances of the highest-frequency phonons in bulk NdNiO$_3$ (green) and LaAlO$_3$ (blue). (c) Comparison between the vibrationally induced changes at the $(505)_{Pbnm}$ and $(055)_{Pbnm}$ reflections, detected +4ps after the excitation. The solid lines are Gaussian fits to the data.

**Fig. 2:** Static photon energy dependent intensity of the NdNiO$_3$ $(505)_{Pbnm}$ reflection (blue dots), measured at the LCLS free electron laser in the vicinity of the Ni K-edge at 40 K nominal sample temperature. These data are normalized to the x-ray intensity incident on the sample and compared to the $(101)_{Pbnm}$ reflection of a NdNiO$_3$ thin film grown on a SrTiO$_3$ substrate, measured at 7 K using synchrotron light (red solid line, data taken from Ref. 22). (b) Transient peak intensity of the NdNiO$_3$ $(505)_{Pbnm}$ reflection, measured at x-ray energies resonant (8.346 keV, blue data) and off-

resonant (8.329 keV, green data) with the Ni K-edge, induced by the same LaAlO$_3$ phonon excitation with a 3.6-mJ/cm$^{-2}$ fluence. Again, these diffraction intensities are normalized to the incident x-ray intensity. The measured resonant diffraction intensity comprises a charge-order contribution (illustrated by the grey shaded region), which disappears on a time scale shorter than that of the lattice dynamics. (c) Time evolution of the charge order parameter, calculated from the data shown in panel (b) as described in the text.

**Fig. 3.** Transient low-frequency THz reflectivity of a NdNiO$_3$/LaAlO$_3$ heterostructure, induced by the direct lattice excitation of the LaAlO$_3$ substrate at 10 K. The insulating nickelate film is transformed to the metallic state within less than 2 picoseconds.

**Fig. 4:** (a) Short time scale spatiotemporal dynamics of the NdNiO$_3$ lattice, magnetic and insulator-metal dynamics along the thin film out-of-plane direction. The structural phase front (black dots, extracted from data presented in Fig. 1(b)) propagates from the interface across the ~13 nm film at about the bulk speed of sound (4.1 nm/ps, grey solid line). The red data points represent the magnetic melt front, copied from Reference 8. The insulator-metal front induced by the same excitation in a ~33 nm NdNiO$_3$ film (extracted from the THz data shown in Fig. 3) is shown as the blue data points, together with the charge order melt front (blue solid line, extracted from the resonant diffraction data presented in Fig. 2(c)). The grey shaded area indicates that data were taken on samples of different thicknesses. (b) Illustration of the involved phase front propagation. At equilibrium, the material is antiferromagnetically and charge ordered. Following excitation of the substrate lattice, loss of charge disproportionation, magnetic order melting and the lattice rearrangement advance at different speeds across the film.

**Figure 1**

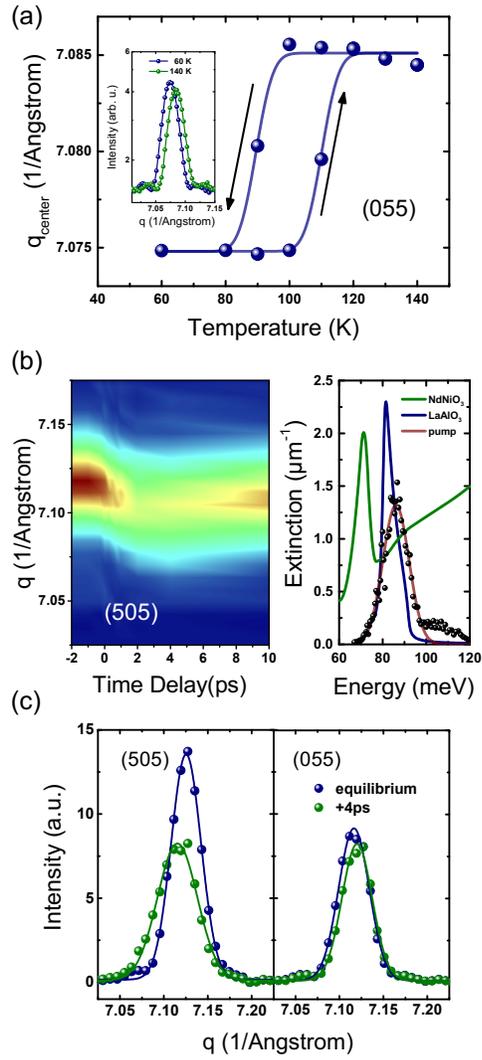

**Figure 2**

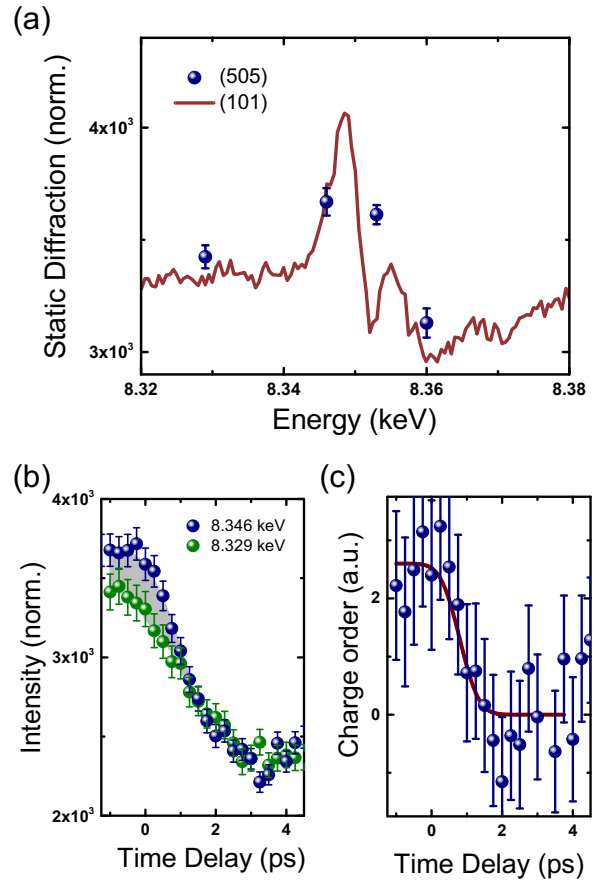

**Figure 3**

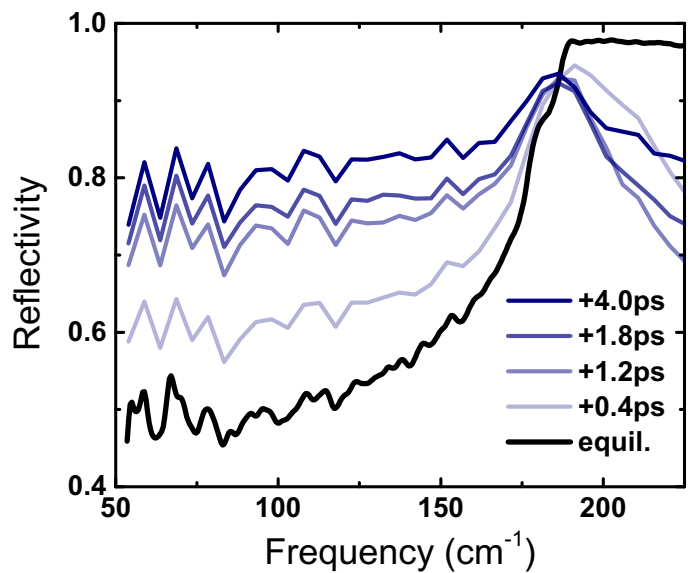

**Figure 4**

(a)
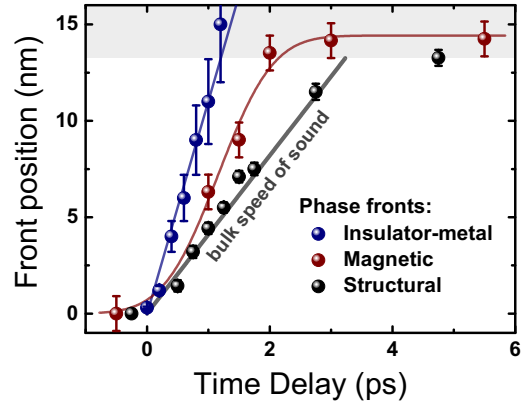

(b)
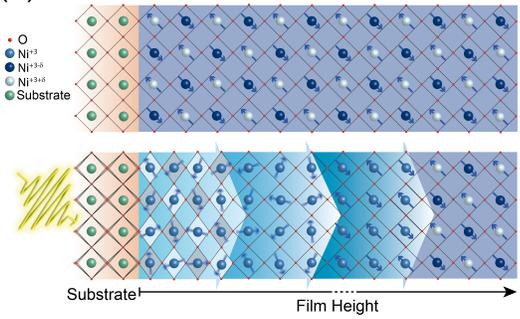